\documentclass[12pt,a4paper]{article}

\usepackage{amsmath,amsfonts,amssymb,bm}

\usepackage{graphicx} 

\usepackage[hang,normalsize,bf]{subfigure} 

\begin{document}

\begin{center}

{\Large \bf The Spectator Electromagnetic Effect \\
on Charged Pion Spectra\\ 
in Peripheral
Ultrarelativistic \\
Heavy-Ion Collisions\\
}

\vspace {0.6cm}

{\large Andrzej Rybicki $^{1}$}

\vspace {0.3cm}

{\large and}

\vspace {0.3cm}

{\large Antoni Szczurek $^{1,2}$}

\vspace {0.2cm}

$^{1}$ {\em Henryk Niewodnicza\'nski Institute of Nuclear Physics, 
Polish Academy of Sciences,
PL-31-342 Krak\'ow, Poland\\}
$^{2}$ {\em University of Rzesz\'ow\\
PL-35-959 Rzesz\'ow, Poland\\}

\end{center}

\begin{abstract}

We estimate the electromagnetic effect of the spectator charge on the
momentum spectra of $\pi^+$ and $\pi^-$ produced in peripheral Pb+Pb
collisions at SPS energies. We find that the effect is large and results
in strongly varying structures in the $x_F$ dependence of the
$\pi^+/\pi^-$ ratio, especially at low transverse momenta where a 
deep valley in the above ratio is predicted at $x_F \sim$ 
0.15 -- 0.20.

It appears that the effect depends on initial conditions. Thus, it
provides new information on the space and time evolution of the 
non-perturbative pion creation process.
\end{abstract}


\section{Motivation}

The process of pion production in high energy hadronic reactions (like 
p+p, p+nucleus or nucleus+nucleus collisions) belongs to the domain of the 
strong interaction theory: quantum chromodynamics (QCD). For an enormous 
majority of pions produced, the mechanism underlying this process 
 is non-perturbative. Here, QCD
 does not lead to 
quantitative predictions. Instead, phenomenological models are 
used~\cite{kittel}. These contain arbitrary scenarios of the hadronic 
collision dynamics, and evidently rely on experimental input to 
differentiate right scenarios from wrong ones.

Most of the experimental input collected so far (at ISR, SPS, RHIC, and
others) provides exclusively information about final state momentum space
($p_x$, $p_y$, $p_z$). Observables such as inclusive distributions of
kinematical variables, or their correlations, give no information on how
the reaction occurs in position space ($x$,$y$,$z$) or in time ($t$). To
get such information, a specific phenomenon directly dependent on the
reaction evolution in space and time has to be isolated. The HBT
effect~\cite{baym} can be quoted as an example of such a phenomenon.

This paper considers another possible candidate for such a phenomenon in
ultrarelativistic heavy-ion collisions. We mean here the electromagnetic
interaction between the remnants of the two nuclei (spectator systems) and
the positive and negative pions produced in the course of the collision.  
The two highly charged spectator systems moving at relativistic velocities
generate a rapidly changing electromagnetic field which modifies the
original pion trajectories. This causes a distorsion of 
observed kinematical pion spectra.  
It is to be expected that this
distorsion is interrelated to the dynamics of the collision, and
in particular to the time evolution and initial conditions of the
participant and spectator zones.

Our aim is to study this electromagnetic ``Coulomb'' effect for the
specific case of peripheral Pb+Pb collisions at SPS energies (158
GeV/nucleon beam energy, $\sqrt{s}_{NN}$=17 GeV). 
 We choose peripheral collisions because they are characterised by the 
largest spectator charge.
 Our specific tasks are:

 \begin{enumerate}
 \item[1.] to provide a general description of main features of the effect 
in the framework of a simplified but realistic, relativistic two-spectator 
model.
 \item[2.] to define the regions of produced pion phase-space that are 
particularly sensitive to the electromagnetic influence of the two 
spectator systems.
 \item[3.] to get a first idea on whether the electromagnetic effect 
depends on the evolution of the pion production process in space and time. 
\end{enumerate}

Seen in a more general context, our analysis is complementary to various 
studies of the role of specific phenomena like multiple 
collisions~\cite{mul}, isospin effects~\cite{isosp}, the neutron 
halo~\cite{PS04}, or Fermi motion~\cite{SB04} in particle production in 
heavy-ion reactions. To understand the role played by all these effects 
seems to us a necessity if a realisitic description of the 
non-perturbative heavy-ion collision dynamics is to be provided.

The remainder of this paper is organized as follows. The general context
of Coulomb interactions in nuclear collisions is shortly discussed in
Section~2. The general idea and the details of our model are described in
Section~3. The results of our study are presented in Section~4.  
Conclusions are formulated in Section~5.

\section{Coulomb Effects at Low and
High Energies}

Various types of Coulomb effects were investigated both experimentally and 
theoretically in the past; some examples will be enumerated below.

Various results are available in the low (or intermediate) energy regime,
up to a few GeV/nucleon. On the experimental side, it was realized in
studies at Bevalac that the Coulumbic field causes extremely large effects
at forward angles, i.e. at projectile velocity \cite{Bevalac1,Bevalac2}.  
The Coulombic effect was also observed at subthresold energies
\cite{CERN_low_energy}. In studies of p+Au and He+Au reactions at several
GeV/nucleon, the non-relativistic approach to the Coulomb field brought
precise information on the evolution of the nuclear fragmentation
process~\cite{karnaukhov}. Finally, an interesting charge asymmetry was
found in the kinetic energy spectrum of particles produced in collisions
of 9 GeV protons with emulsion nuclei~\cite{fried}.

The situation on the theoretical side is rather complicated. Attempts were 
made to describe the intermediate energy minimum bias data by a 
multiparametric modelling of the pion source fireballs, and calculating 
the subsequent transport of pions in the Coulomb field of the 
spectators~\cite{LK79,CK81}. Quantal effects were discussed 
in~\cite{GK81}. Other simple models were considered in 
\cite{BB87,L95,BBGH98}.

In the high energy regime, studies of spectator Coulomb interactions offer 
two advantages relative to the low energies. On the experimental side, 
impact parameter (centrality) selection becomes possible by produced 
particle multiplicity measurement or by forward calorimetry. This allows a 
selection of peripheral collisions where the spectator effect will be 
the largest. On the theoretical side, seen in the context of typical 
relativistic pion velocities, the break-up of the spectator system can be 
assumed to be slow. This simplifies the picture. However, a 
fully relativistic description of electromagnetic effects involving e.g. 
retardation phenomena becomes a necessity at high energies.

In early cosmic ray experiments, large $\pi^+/\pi^-$ asymmetries were 
observed for produced slow pions~\cite{yagoda}. As far as nucleus+nucleus 
collisions and SPS energies are concerned, various 
experimental~\cite{NA44,E866,wa98} and theoretical~\cite{OSBW96} papers on 
the influence of the electromagnetic field on particle production in the 
mid-rapidity ($x_F=~0$) region\footnote{Note:
 the Feynman variable $x_F=2p_L/\sqrt{s}$, the transverse momentum $p_T$, 
the energy $E$, and other related quantitites will always be considered in 
the nucleon+nucleon c.m.s.}
 of Pb+Pb reactions are known. However, these focus on the
role played by the electromagnetic field originating from particles taking
direct part in the reaction.

As far as the influence of the two spectator systems in the rest of
available phase-space is concerned, we are aware of only one published
measurement~\cite{NA52}. This has been performed for a very limited
acceptance range (forward angles i.e.  $p_T=0$). 
 New, better measurements can be expected in future from the NA49
apparatus~\cite{prel}.
 No theoretical description is known to us. To
provide such a simplified but realistic description is the main purpose of
the present paper.

\section{Modelling a Peripheral Pb+Pb Collision at SPS Energies}

\begin{figure}[htb] 
\begin{center}
\includegraphics[width=10cm]{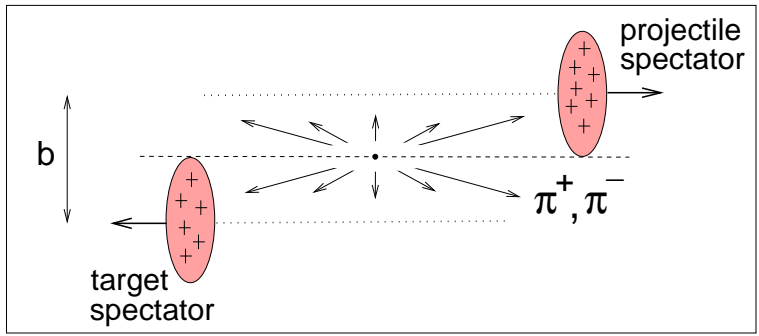}
 \caption{\it Our simplified view of a Pb+Pb collision. The hypothetical
pion emission region is reduced to a single point in position space (see
text). 
 \label{ideowy}
}
\end{center}
\end{figure}

Our aim is to obtain a realistic estimate of the spectator-induced 
electromagnetic effect, avoiding however
the discussion of complex and poorly known details of soft hadronic
particle production. Thus we decide on a maximally simplified approach,
illustrated in Fig.~\ref{ideowy}. A
peripheral Pb+Pb reaction can be imagined to consist of three steps:

 \begin{enumerate}
 \item
 the collision takes place at a given impact parameter $b$. The two highly 
charged spectator systems follow their initial path with essentially 
unchanged momenta;
 \item
 the participating system evolves until it finally gives birth to final 
state pions. The evolution of the pion emission region in space and 
time is {\em a priori} unknown;
 \item
 charged pion trajectories are modified by electromagnetic interaction 
with the spectator charge; the spectator systems undergo a complicated 
nuclear fragmentation process.
 \end{enumerate}

We model these steps in the following simplified way:

\begin{enumerate}
 \item
 a peripheral Pb+Pb collision involving 60 
participating nucleons is assumed. This corresponds to an impact 
parameter 
$b$ equal to 10.61~fm (see Sec.~\ref{colgeo}).
 The two spectator systems are modelled as two uniform spheres in their
respective rest frames. The sphere density is the standard nuclear
density $\rho=0.17$/fm$^{3}$. The total positive charge of each spectator
system is $Q=70$ elementary units. In the collision rest frame the two
spheres become charged disks (Fig.~\ref{ideowy}).
 \item
 the pion emission region is reduced to a single point in space, namely
the original interaction point. The emission time $t_E$ is a free
parameter in our model. For peripheral Pb+Pb reactions studied here, we
assume that the initial two-dimensional ($x_F,p_T$) distribution of the
emitted pion is similar to that in underlying nucleon+nucleon collisions.
 \item
 charged pions, with their initial momentum vector defined above in 
point 2., are 
numerically traced in the electromagnetic field of the spectator 
charges until they reach a distance of 10,000 fm away from the original 
interaction point and from each of the two spectator systems. The 
fragmentation of the spectator systems is neglected;  the influence of 
participant 
charge, strong final state interactions, etc, are not considered.
 \end{enumerate}

For each of the above steps, the details of our approach are explained
below.


\subsection{Collision Geometry}
\label{colgeo}

\begin{figure} 
\begin{center}
\includegraphics[width=5cm]{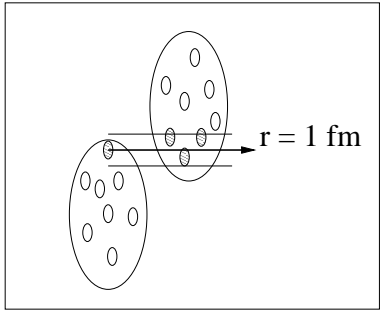}
\caption{\it Geometrical model of a Pb+Pb collision. The filled 
ellipses represent participating nucleons while empty ellipses represent 
spectator nucleons.}
\label{geom}
 \end{center}
 \end{figure} 

We adjust the geometry (centrality) of the considered Pb+Pb collision to 
60 participating nucleons in order to make it comparable to that of 
peripheral heavy-ion data samples collected at the SPS~\cite{ferenc}.

The relation between the reaction impact parameter $b$, the number of 
participating nucleons $N_{part}$ and the spectator charge $Q$ is defined 
by the nuclear density profile and the elementary nucleon+nucleon 
cross section. We study this relation by means of a geometrical Monte 
Carlo simulation. This simulation (Fig.~\ref{geom}) produces spatial 
distributions of protons and neutrons using nuclear density profiles 
obtained for $^\mathrm{208}$Pb by Mizutori et al. in the 
Hartree-Fock-Bogoliubov (HFB)  approach~\cite{dobaczewski-wxyz}. For each 
Pb+Pb reaction, a nucleon is defined as participant if it is crossed by 
one or more nucleons from the adverse nucleus within a transverse radius 
of less than 1 fm. This corresponds to an elementary nucleon+nucleon 
cross-section of 31.4 mb in good agreement with experimental 
data~\cite{pp}. No re-interaction with the spectator system is considered.

Technically, the simulation is performed at a fixed value of the impact 
parameter $b$. We obtain the average number of participants 
$N_{part}=60$ at $b=10.61$~fm. At the same time we 
obtain\footnote{The
 spectator charge $Q$ is a numerical result of the simulation. It does not 
{\em a priori} follow from $N_{part}$ due to possible neutron halo 
effects~\cite{Trzcinska01}. Our Monte-Carlo takes account of these 
effects.}
 the average charge of the spectator system $Q\approx70$ e.u.

\begin{figure} 
\begin{center}
\includegraphics[width=5cm]{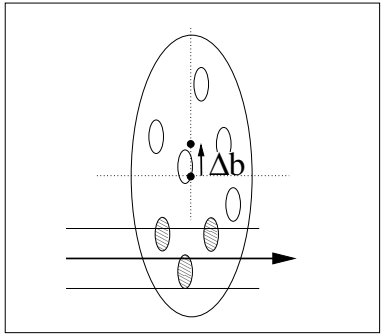}
 \caption{\it As some of the nucleons become participants, the center of 
gravity of the spectator protons is displaced relative to the center of 
gravity of the original Pb nucleus.
 \label{geomr}}
\end{center}
 \end{figure}

Another important geometry parameter to be estimated is the displacement 
$\Delta b$ of the spectator protons' center of gravity relative to the 
center of gravity of the original Pb nucleus. This displacement originates 
from trivial geometrical reasons (Fig.~\ref{geomr}). Our simulation gives 
$\Delta b=0.76$~fm. Thus the effective distance of closest approach 
between the centers of gravity of the two spectator systems will be 
$b^{'}_\mathrm{}=b+2\Delta b=12.13$~fm.

We consider that the exact shape of the spectator systems is not essential 
for our subsequent simulations of electromagnetic effects. For the sake of 
clarity, we model the two spectator systems as two homogenous spheres with 
standard nuclear densities $\rho$=0.17/fm$^3$ and with properly shifted 
centers of gravity (Fig.~\ref{ideowy}). As discussed above, the distance 
of closest approach of the two spheres' centers of gravity is 
$b^{'}_\mathrm{}=12.13$~fm. The total charge of each sphere is $Q=70$ 
elementary units.

For the case of peripheral collisions considered here, estimates on
geometry parameters may strongly rely on a proper description of the
nuclear profile. Therefore, in order to cross-check our results, we
performed an independent simulation using a different nuclear profile: a
phenomenological parametrization obtained by Trzci\'nska et al. on the
basis of antiprotonic atom data~\cite{atph}. With respect to the Mizutori
profile, the deviations on $b^{'}_\mathrm{}$, $Q$ and $N_{part}$ are small
and remain below 1.5\%. Thus we consider our estimates as fairly precise.


\subsection{Initial Pion Emission}
\label{emission}

Having established the initial geometry of the peripheral Pb+Pb reaction, 
we subsequently model the emission of produced $\pi^+$ and 
$\pi^-$.

 We reduce the unknown initial emission region to a unique point in space, 
namely the original interaction point (Fig.~\ref{ideowy}). We assume one 
emission time $t_E$ which is a free parameter of our model; various values 
of $t_E$ will be considered. Such a simplification of initial conditions 
gives a convenient way to estimate the dependence of the electromagnetic 
effect on the characteristics of the initial pion emission process (like 
the pion formation time or its distance from the two spectator systems).

We assume that the initial kinematical spectra of the emitted pions are 
similar to these in nucleon+nucleon collisions and that they 
follow wounded nucleon scaling~\cite{wnm}. Full azimuthal symmetry of the 
emission is assumed. The detailed shape of the emitted pion 
density distribution requires an additional discussion.

A very precise and complete set of experimental data on $\pi^+$ and 
$\pi^-$ production in p+p collisions is available at the SPS~\cite{pp}. 
The published double differential spectra of pions in $x_F$ and $p_T$ 
display a strong dependence on pion isospin. They also contain local shape 
structures attributed to production of hadronic resonances and their 
subsequent decay into pions. It has been demonstrated that such elementary 
effects reappear in nuclear collisions~\cite{isosp}. Any attempt at an 
in-depth description of Pb+Pb interactions must take them into account.

Such a description is, however, beyond our specific scope. As we 
concentrate on the spectator-induced Coulomb effect {\em per se}, we need 
a clear distinction between hadronic and electromagnetic phenomena to 
allow for an easy interpretation of our results. Thus, we simplify the 
situation:

\begin{enumerate}
 \item we neglect isospin effects i.e. assume equal initial emission
spectra for $\pi^+$ and $\pi^-$;
 \item we construct a smooth two-dimensionnal shape that reproduces the 
most basic features of the pion production data~\cite{pp}; it does not 
include the more subtle, local shape structures.
 \end{enumerate}

\begin{figure}[t] 
\vspace*{-10cm}
\begin{center}
\hspace*{-5.7cm}\includegraphics[width=13.6cm]{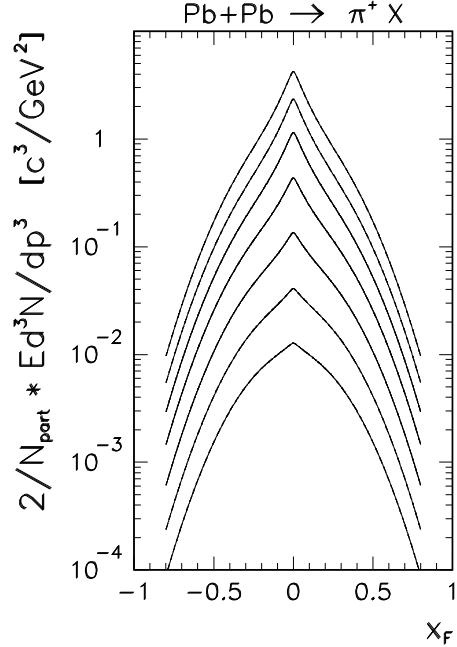}
 \caption{\it Assumed invariant density of $\pi^+$ emitted per 
participant pair in
Pb+Pb collisions, drawn as function of $x_F$ at fixed $p_T$
values. From top to bottom, the curves correspond to $p_T=$ 50, 100,
200, 400, 600, 800, and 1000 MeV/c. For clarity, the two top
curves are multiplied by 2.5 and 1.5, respectively. The invariant density 
of emitted $\pi^-$ is assumed to be identical to the above (see text).
}
 \label{parametrization}
 \end{center}
 \end{figure} 

The resulting two-dimensional invariant pion density\footnote{Note
 that in terms of $x_F$, $p_T$ and azimuthal angle $\phi$, our invariant 
pion density per event and per participant pair writes
 $\frac{2}{N_{part}}~E\frac{d^3N}{dp^3}
 =\frac{2}{N_{part}}\frac{2E}{p_T\sqrt{s}}\frac{d^3N}{dx_Fdp_Td\phi}$ \; .}
 is drawn in Fig.~\ref{parametrization}. We do not consider values 
higher than $|x_F|=0.8$. We checked that below $p_T=1$ GeV/c, our curves 
typically deviate from measured average pion ($\frac{\pi^++\pi^-}{2}$) 
densities by about 10\%. Maximal deviations reach up to 21\% of our 
assumed density.  Such a gross description is sufficient for our specific 
aim. Evidently it cannot be used as a ``fit'' nor ``parametrization'' of 
the data. Note that the failure of oversimplified analytical 
parametrizations to reproduce the p+p data has been demonstrated 
in~\cite{pp}.

For completeness, we write the analytical form of our 
emitted pion density per participant pair:

\begin{equation}
\frac{2}{N_{part}}~E\frac{d^3N}{dp^3}_{Pb+Pb\rightarrow\pi X} 
=
\sum_{n=1,2} a_n\exp\left( -(x/b_n)^{c_n} \right)
\exp\left( -u_T/d_n \right) \; ,
\label{param}
\end{equation}
where   $\pi=\pi^+$ or $\pi^-$, $N_{part}=60$ (Sec.~\ref{colgeo}), $x = 
\sqrt{ x_F^2 + g^2 }$, $u_T = \sqrt{q^2 + p_T^2}$,
and numerical parameter values are listed below:
\begin{center}
\begin{tabular}{  c | c  c  c  c | c l }
$n$ & $a_n$ [c$^3$/GeV$^2$] & $b_n$ & $c_n$ & $d_n$ [GeV/c]  \\
\hline
1   & 2.32229 & 0.369967 & 2. &  0.191506 & $g=$ &    0.01\\
2   & 24.4563 & 0.0873833 & 1.001 & 0.12  & $q=$ & 0.334968 
[GeV/c]\\
\end{tabular}
\end{center}

\subsection{Propagation of Charged Pions in
the Spectator Electromagnetic Field}
\label{prop}

The initially produced charged pions are subjected to the electromagnetic 
field of the two spectator systems moving at relativistic velocities.  

We choose the overall center of mass system of the collision to calculate
the evolution of pion trajectories. For symmetric Pb+Pb collisions and
neglecting the Fermi motion effect (see \cite{SB04}) this is also the
nucleon+nucleon center of mass system.

We assume that for ultrarelativistic Pb+Pb reactions, the nuclear 
fragmentation process is slow relative to the relativistic pion velocities 
and can be neglected in a first approximation. Thus the spectator nucleon 
velocity remains constant and identical to the velocity of the parent Pb 
ion.

We define our space coordinate system in the way shown in Fig.~\ref{coor}.  
For simplicity, we label the two spectator systems as left ($L$) and right
($R$).  We set the time scale such that at $t=0$ the center of gravity of
each of the spectator systems is found at $z_L = z_R = 0$. Thus the
time-dependent position of the moving spectators is:

\begin{figure} 
\begin{center}
\includegraphics[width=10cm]{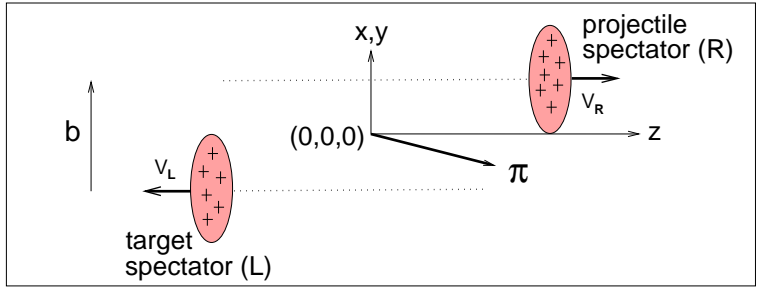}
 \caption{\it The space coordinate system for the Pb+Pb collision. The 
origin $(x,y,z)=(0,0,0)$ is located at the interaction point. The 
projectile and target spectators are denoted left (L) and right (R). We 
define the impact parameter $\vec{b}$ as a vector.
 \label{coor}
}
\end{center}
\end{figure}
 %
%
%
 \begin{eqnarray}
 \vec{R}_{L}(t) &=& -\vec{b}/2 + \vec{v}_{L}\cdot t \; , \nonumber \\
 \vec{R}_{R}(t) &=& \;\;\; \vec{b}/2 + \vec{v}_{R}\cdot t \; . 
 \end{eqnarray}
 Here, $\vec{v}_{L}$ and $\vec{v}_{R}$ are the two spectator velocity 
vectors. For the SPS energy discussed here $v_{L}=v_{R}\equiv v_S=0.994c$. 
The value of the impact parameter vector $\vec{b}$ is set to $b=12.13$~fm 
(see Sec.~\ref{colgeo}).

Each of the two spectator systems defines its own rest frame. We define 
$\vec{E}'_{L}$ as the constant electrostatic field generated by the 
spectator $L$ in its rest frame, and $\vec{E}''_{R}$ as the field 
generated by the spectator $R$ in its rest frame. Having assumed both 
systems as uniform spheres with a normal nuclear density 
$\rho=0.17$/fm$^{3}$ and with a total charge $Q=70$ elementary units 
(Sec.~\ref{colgeo}) we write:
\begin{equation}
  \vec{E}'_L(\vec{r'_c}) = 
  \begin{cases}
    &k \; Q \;\; \vec{r'_c}  \; / \; {r'_c}^{3} \text{~~~for~~~} r'_c > 
R_s \\
    &k \; Q \;\; \vec{r'_c}  \; / \; R_{s}^{\;3} \text{~~~for~~~} 
r'_c < R_s
  \end{cases}
  \label{static_E_L}
 \end{equation}
%
 \begin{equation}
  \vec{E}''_R(\vec{r''_c}) =
  \begin{cases} 
    &k \; Q \;\; \vec{r''_c} / {r''_{c}}^{\;3} \text{~~~for~~~} 
r''_{c} > R_s \\
    &k \; Q \;\; \vec{r''_c} / R_{s}^{\;3} \text{~~~~for~~~} r''_{c} < 
R_s
  \end{cases}
 \label{static_E_R}
 \end{equation}
 %
%
 In the equations above, $\vec{r'_c}$ ($\vec{r''_c}$) is the position
relative to the center of the $L$ ($R$) spectator, defined in this
spectator's rest frame. $k\approx 1.44~\text{MeV}\cdot\text{fm}/e^2$ is the electrostatic constant.
 $R_s = [N_{spec} / (4/3 \pi \rho)]^{1/3}$ is the sphere radius defined by 
the number of spectator nucleons $N_{spec}$. For the case considered here 
$R_s=6.3$~fm.

We transform the fields $\vec{E}'_{L}$,
$\vec{E}''_{R}$ to the center of mass system.
 Here, the moving spectator
 charge generates both electric and magnetic fields.
 From the general Lorentz transformation~\cite{Jackson}
we get
\begin{equation}
\begin{split}
\vec{E}_{L}(\vec{r},t) &= \gamma_s \vec{E}'_{L}
(\vec{r'_c})
 - \frac{\gamma_s^2}{\gamma_s+1}
\; \frac{\vec{v}_{L}}{c} \; \left( \frac{\vec{v}_{L}}{c} \cdot 
\vec{E}'_{L}
(\vec{r'_c})
\right)
\; ,
 \\
\vec{B}_{L}(\vec{r},t) &= \gamma_s \left( \frac{\vec{v}_{L}}{c} \times 
\vec{E}'_{L} 
(\vec{r'_c})
\right)
\end{split}
\end{equation}
for the left spectator and
\begin{equation}
\begin{split}
\vec{E}_{R}(\vec{r},t) &= \gamma_s \vec{E}''_{R}
(\vec{r''_c})
 - \frac{\gamma_s^2}{\gamma_s+1}
\; \frac{\vec{v}_{R}}{c} \; \left( \frac{\vec{v}_{R}}{c} \cdot 
\vec{E}''_{R}
(\vec{r''_c})
\right)
\; ,
\\
\vec{B}_{R}(\vec{r},t) &= \gamma_s \left( \frac{\vec{v}_{R}}{c} \times 
\vec{E}''_{R} 
(\vec{r''_c})
\right)
\end{split}
\end{equation}
 for the right spectator. 

In the equations above, the $\gamma_s$ factor is defined as 
$\gamma_s=(1-v^2_s/c^2)^{-1/2}$. The vectors $\vec{E}_{L}$ $(\vec{E}_{R})$ 
and $\vec{B}_{L}$ $(\vec{B}_{R})$ are respectively the electric and 
magnetic fields generated by the left (right) spectator at the space-time 
position $(\vec{r},t)$. The relation between $\vec{r}=(x,y,z)$ and the 
spectator rest frame coordinates $\vec{r'_c}=(x'_c,y'_c,z'_c)$, 
$\vec{r''_c}=(x''_c,y''_c,z''_c)$ is given by the Lorentz transformation 
and the impact parameter vector $\vec{b}=(b_x,b_y,0)$:
\begin{equation}
  \begin{cases}
    & x'_c=x+b_x/2 \\
    & y'_c=y+b_y/2 \\
    & z'_c=\gamma_s(z+v_s t)
  \end{cases}
~~~~~~~~~~~~
%
  \begin{cases}
    & x''_c=x-b_x/2 \\
    & y''_c=y-b_y/2 \\
    & z''_c=\gamma_s(z-v_s t)
  \end{cases}
\label{essential_distance}
 \end{equation}

 We now consider a charged pion emitted at time $t=t_E$ from the 
interaction point $\vec{r}=(0,0,0)$ with its initial momentum 
$\vec{p}_\pi(t=t_E)$ specified by Eq.~(\ref{param}) from 
Sec.~\ref{emission}. This situation is illustrated in Fig.~\ref{coor}. 
The Lorentz force acting on the pion is:
\begin{equation}
\frac{d \vec{p}_{\pi}}{dt} = \vec{F}_{\pi}(\vec{r},t) =
 q_{\pi} \left( \vec{E}(\vec{r},t) 
+ \frac{\vec{v}_{\pi}(\vec{r},t)}{c} \times \vec{B}(\vec{r},t) \right) \; ,
\label{Lorentz_force}
\end{equation}
 where $q_\pi$ is the pion charge, $\vec{E}(\vec{r},t) = 
\vec{E}_{L}(\vec{r},t) + \vec{E}_{R}(\vec{r},t)$ and $\vec{B}(\vec{r},t) = 
\vec{B}_{L}(\vec{r},t) + \vec{B}_{R}(\vec{r},t)$ are standard 
superpositions of fields from the two sources.

 The resulting pion trajectory $\vec{r}_{\pi}(t)$ is 
defined by its time-dependent velocity $\vec{v}_\pi(\vec{r},t)$:
\begin{equation}
\frac{d \vec{r}_{\pi}}{d t} = \vec{v}_{\pi}(\vec{r},t) = 
\frac{\vec{p}_\pi \; c^2}{\sqrt{p^2_\pi+m^2_\pi }} \; ,
\label{velocity}
\end{equation}
where $m_\pi$ is the pion mass.


An important feature of Eq.~(\ref{Lorentz_force}) is that it implicitly 
takes account of relativistic retardation effects. This feature has been 
explicitly demonstrated in~\cite{Jackson}. \\

Technically, the propagation of the pion is made by means of an iterative 
Monte-Carlo procedure. This procedure starts at $\vec{r}=(0,0,0)$ and 
$t=t_E$ and calculates the Lorentz force $\vec{F}_{\pi}(\vec{r},t)$ and 
the corresponding change of pion momentum and position in small steps in 
time. The variable step size depends on the actual distance of the pion 
from the nearest spectator system.
 The procedure is iterated numerically until the distance of the pion from 
the origin $(0,0,0)$ is $r > R_{max}$ and at the same time, the distances 
of the pion from the spectators in their respective rest frames are 
simultanously $r'_{c} > R_{max}$ and $r''_{c} > R_{max}$. We found that 
for geometries considered in the present paper, the value of the parameter 
$R_{max}$ = 10,000 fm is sufficiently large to reproduce asymptotic 
momenta. The procedure is {\em weighted}, that is, each pion is generated 
with its proper weight $\frac{d^2N}{dx_Fdp_T}$ deduced from 
Eq.~(\ref{param}). After the propagation procedure is finished, the same 
weight is used to fill the final state pion spectra. Negatively charged 
pions that do not escape from the spectator potential well are rejected by 
our procedure and do not enter into the final state distribution.


\section{Results}

In this Section the results of our Monte-Carlo studies are discussed. 

\subsection{Charged Pion Spectra}
\label{cps} 

\begin{figure}               
\begin{center}
\includegraphics[height=16cm]{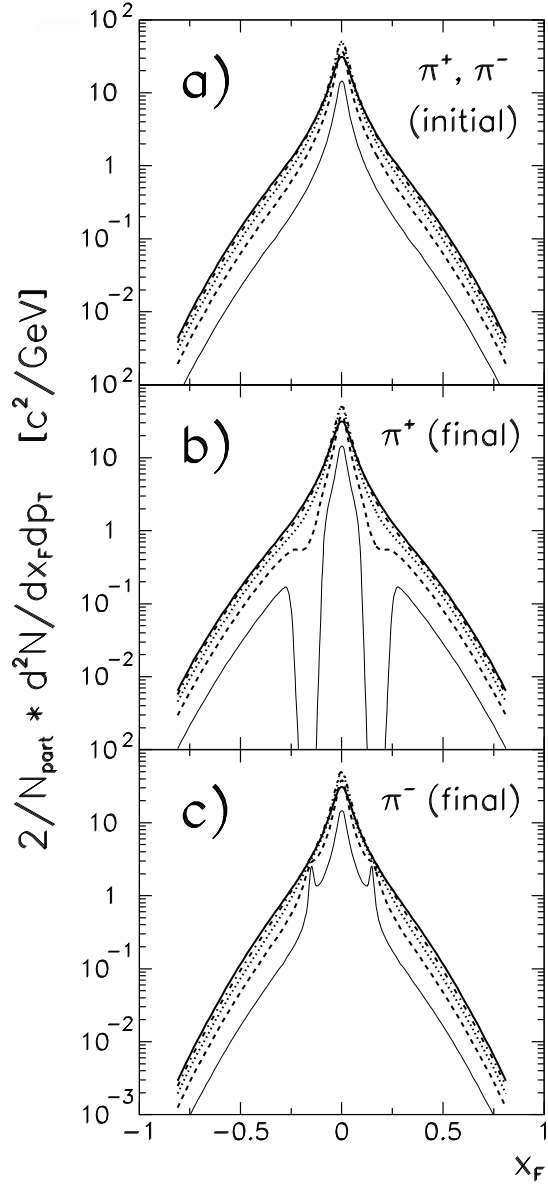}
 \caption{Double-differential density of positively and negatively
charged pions, produced per participant pair in peripheral
Pb+Pb reactions.
 {\bf a)}~Initial density of emitted $\pi^+$ and $\pi^-$.
 {\bf b)}~Density of $\pi^+$ in the final state of the reaction.
 {\bf c)}~Density of $\pi^-$ in the final state of the reaction.  In all
the panels, the pion density is drawn as a function of $x_F$ at $p_T=25$
MeV/c (thin solid), 75 MeV/c (dash), 125 MeV/c (dot), 175 MeV/c
(dash-dot), and 325 MeV/c (thick solid); each of these $p_T$ values
corresponds to a bin of $\pm 25$~MeV/c.  This simulation was made
assuming the pion emission time $t_E$ equal to zero. At negative
$x_F$ reflected curves are drawn.
 \label{dndx}
}
\end{center}
\end{figure}

 We start by estimating the influence which the spectator charge exerts
on double-differential spectra of pions produced in peripheral Pb+Pb
reactions. For the sake of clarity, this part of the discussion remains
limited to the simplest situation where the pion emission time $t_E$ is
equal to zero (immediate pion creation). This description will be
generalized in Sec.~\ref{depini} by considering different
$t_E$ values.

 The situation is illustrated in Fig.~\ref{dndx}. Panel {\bf (a)} shows
the initial spectra of emitted pions. As explained in Sec.~\ref{emission},
in our simple model these spectra are identical for $\pi^+$ and $\pi^-$.  
The presented $\frac{d^2N}{dx_Fdp_T}$ density distributions (scaled down
by the number of participant pairs) follow directly from 
Eq.~(\ref{param}).

In panel {\bf (b)}, the corresponding distributions of $\pi^+$ in the
{final state} of the Pb+Pb reaction are shown. These are obtained by our
Monte-Carlo simulation described in Sec.~\ref{prop}. It is clearly
apparent that the distributions are distorted by the Coulomb repulsion
between the pion and spectator charges. The effect is largest for pions
moving close to spectator velocities ($x_F\approx\pm 0.15$) and at low
transverse momenta ($p_T=25$~MeV/c). Here, two deep valleys in the $\pi^+$
density are visible. A similar but smaller distorsion is also apparent at
$p_T=75$~MeV/c.

An opposite distorsion is present for $\pi^-$ densities shown in panel 
{\bf (c)}. Negative pions are attracted by the positive spectator charge 
and gather at low transverse momenta close to spectator 
velocities. This results in the presence of two large peaks at 
$x_F\approx\pm 0.15$. Remnants of these peaks are apparent at 
$p_T=75$~MeV/c.

Thus, the Coulomb field induced by the spectator charge at SPS energies 
appears strong enough to produce visible distorsions on the pion 
$\frac{d^2N}{dx_Fdp_T}$ distribution, which extends over several orders of 
magnitude in terms of pion density and over many GeV in terms of kinetic 
energy\footnote{In this context, it is interesting to note that the 
initial spectator c.m.s. electrostatic potentials induced on the pion are 
relatively small ($\frac{kQq}{R}\approx 17$~MeV for the case considered in 
Fig.~\ref{dndx}).}.


\subsection{${\pi^+/\pi^-}$ Ratios}
\label{prat} 

Further, more precise information on the spectator electromagnetic effect 
can be obtained by considering the density ratios of produced positive 
over negative pions: ${\pi^+/\pi^-}$. In our model, the ${\pi^+/\pi^-}$ 
ratio for {initially emitted} pions is by definition equal to unity in the 
whole phase space (Sec.~\ref{emission}).  Any deviation of this ratio from 
unity in the final state of the Pb+Pb reaction is therefore directly 
resulting from the spectator-induced Coulomb interaction.

\begin{figure}             
\begin{center}
\includegraphics[height=17cm]{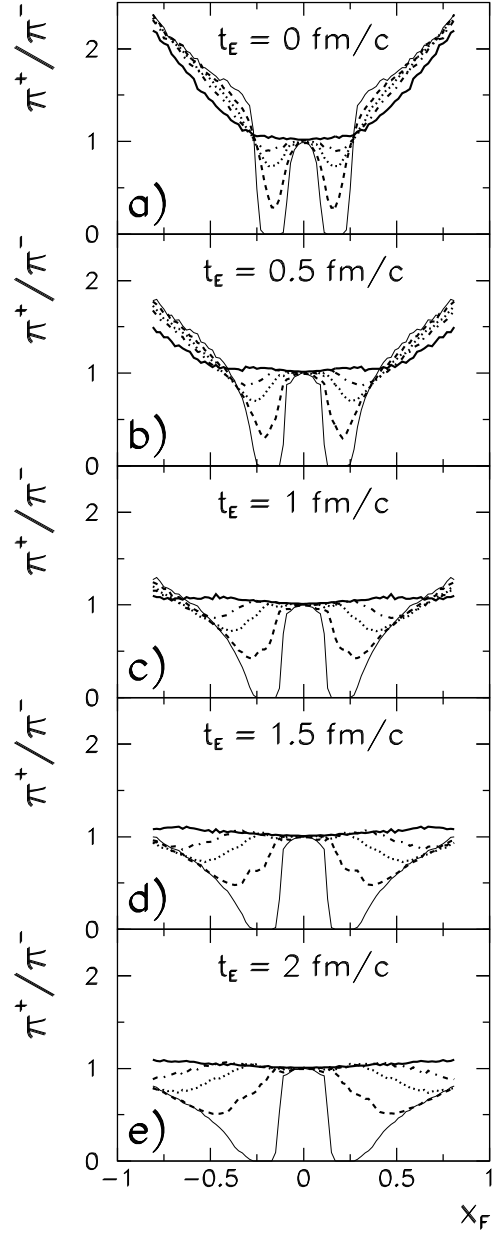}
 \caption{Ratio of density of produced $\pi^+$ over produced $\pi^-$ in 
the final state of the peripheral Pb+Pb reaction, obtained for five values 
of the pion emission time $t_E$. In all the panels, the $\pi^+/\pi^-$ 
ratio is drawn as a function of $x_F$ at $p_T=25$ MeV/c (thin solid), 75 
MeV/c (dash), 125 MeV/c (dot), 175 MeV/c (dash-dot), and 325 
MeV/c (thick solid). The small cusps on the curves correspond to the 
statistical fluctations in our Monte Carlo. At negative $x_F$ reflected 
curves are drawn.
 \label{rat}}
 \end{center}
 \end{figure}

Fig.~\ref{rat}{\bf(a)} shows the $x_F$-dependence of the ${\pi^+/\pi^-}$ 
ratios computed for the case considered in the preceding Section, namely 
the pion emission time $t_E$ assumed equal to zero. The spectator Coulomb 
field appears now to produce a characteristic, complex pattern of 
deviations from unity. The first element of this pattern is a double, 
two-dimensional valley which covers the low-$p_T$ region in the vicinity 
of $x_F\approx\pm 0.15$; the valley remains still visible at 
$p_T=175$~MeV/c. The second element is a smooth rise of the 
${\pi^+/\pi^-}$ ratio at higher $|x_F|$. This rise is present for 
all the considered $p_T$ values; at fixed $x_F$, the ratio slowly 
decreases with increasing $p_T$.

\begin{figure}            
\begin{center}
\includegraphics[height=15cm]{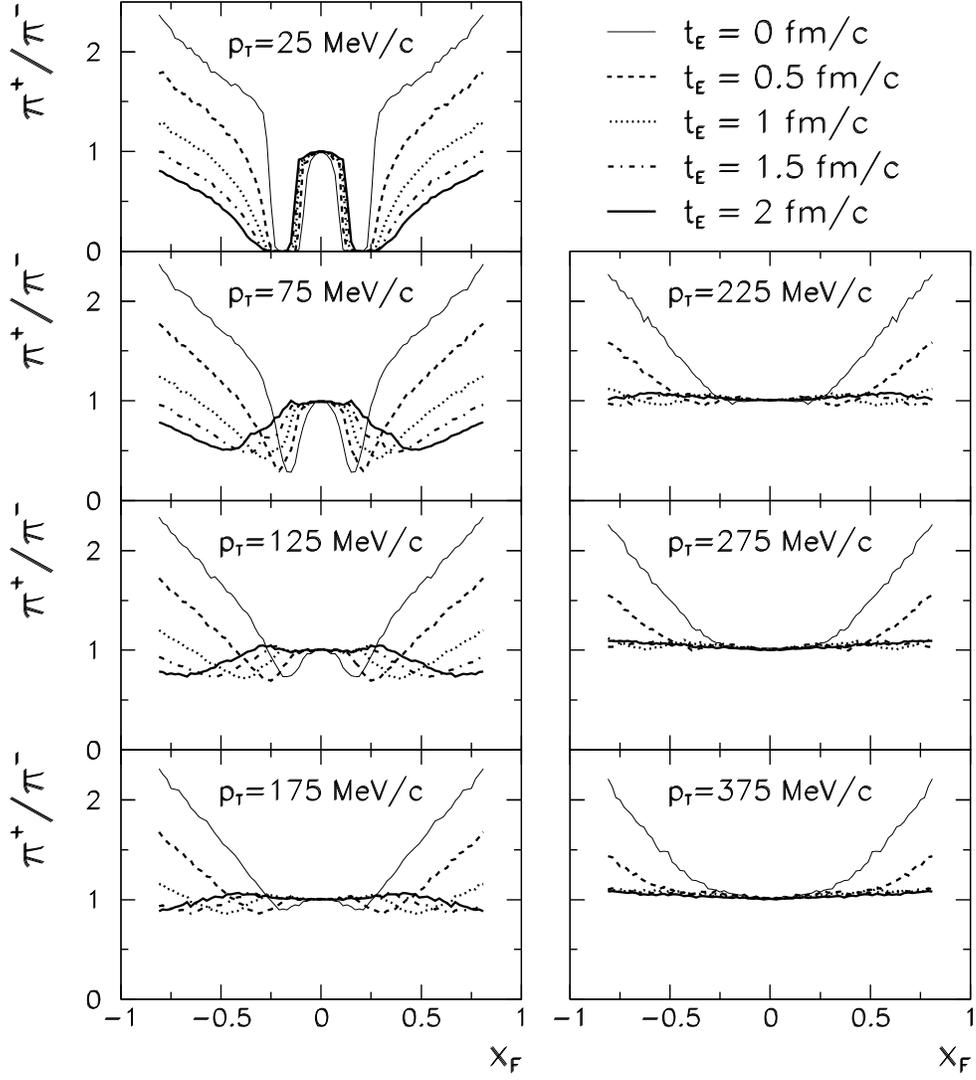}
 \caption{Final state $\pi^+/\pi^-$ ratio for the
peripheral Pb+Pb reaction, shown at fixed values of $p_T$ as a function of
$x_F$. The five considered values of the pion emission time $t_E$ are
differentiated by means of different line types. At negative $x_F$ 
reflected curves are drawn.
 \label{rat-t}}
\end{center}
\end{figure}

\subsection{Dependence on Initial Conditions}
\label{depini}

The central issue of this paper is the sensitivity of spectator-induced 
Coulomb effects to initial conditions of the pion production process.
This issue is studied in Fig.~\ref{rat}{\bf(a-e)} where
the ${\pi^+/\pi^-}$ pattern obtained with the pion emission time
$t_E$ equal to zero is compared to these computed with
$t_E=0.5$, 1, 1.5, and 2~fm/c.

 Indeed, clear differences appear for different $t_E$ values. With 
increasing $t_E$, a broadening of the double valley at $x_F\approx\pm 
0.15$, and a decrease of the ${\pi^+/\pi^-}$ ratio at higher absolute 
$x_F$, are visible.

The study is further quantified in Fig.~\ref{rat-t}. Here, the 
${\pi^+/\pi^-}$ patterns obtained with different $t_E$ values are
directly compared at fixed $p_T$.
It becomes evident that a change of 0.5 fm/c in 
the pion emission time is sufficient to produce visible changes in the 
${\pi^+/\pi^-}$ ratio. In the low transverse momentum region 
($p_T<200$~MeV/c), a displacement of the double valley towards higher 
absolute $x_F$ becomes apparent on top of its broadening with $t_E$. At 
higher transverse momenta, the ratio seems to stabilize for $t_E>1$~fm/c.

This evident sensitivity of the ${\pi^+/\pi^-}$ ratio to the pion
emission time $t_E$ clearly indicates that the Coulomb effect
induced by spectator protons depends on initial conditions imposed
on pion production. Let us 
remind that the different considered values of $t_E$ are, in our model, 
equivalent to different distances between the pion formation zone and the 
two spectator systems. It can therefore be concluded that the Coulomb 
distorsion pattern induced by spectators on $\pi^+$ and $\pi^-$ spectra 
carries information on the evolution of the particle production process 
both in space and in time.

\pagestyle{plain}

\section{Summary and Conclusions}

The electromagnetic interaction between the spectator protons and charged 
pions produced in the peripheral Pb+Pb reaction has been studied by means 
of a simplified model. This interaction is strong enough to produce 
visible distorsions in the final state densities of positive and negative 
pions.

The main feature of this ``Coulomb'' effect is a big dip in the $\pi^+$ 
density distribution at low transverse momenta in the vicinity of 
$x_F\approx\pm 0.15$, accompanied by an increase of $\pi^-$ density
in the  corresponding region of phase space.
This results in the presence of a 
double, two-dimensionnal valley in the $\pi^+/\pi^-$ density ratio. At 
higher absolute $x_F$ another distorsion, namely a smooth increase of 
$\pi^+/\pi^-$ with $x_F$, may appear.

The sensitivity of this electromagnetic effect to initial conditions 
imposed on pion production has been estimated. The effect is clearly 
sensitive to initial conditions. Changes of the pion emission time by 0.5 
fm/c (in c.m.s. time) are sufficient to modify the observed distorsion 
pattern. In our model, such changes are equivalent to changes of position 
of the formation zone by 0.5 fm relative to the two spectator systems. 
Thus, the electromagnetic effect appears to depend on the evolution of 
the pion production process in space and in time.

Our study demonstrates the importance of new, double-differential data on 
the $x_F$ and $p_T$-dependence of pion production in peripheral 
nucleus+nucleus collisions. Further analyses of this subject should take 
account of other effects present in nuclear reactions (multiple 
collisions, isospin, nuclear fragmentation and possibly strong final
state interactions of pions with the spectator systems).
It seems nevertheless clear that the electromagnetic phenomena induced
by the presence of spectator charge carry interesting information on
the mechanism of the non-perturbative particle production process.

\vskip 1cm

{\bf Acknowledgments.}\\
 We wish to express our deep gratitude to Hans Gerhard Fischer for his 
valuable support as well as for numerous fruitful discussions.
We are especially indebted to Andrzej G\'orski for his careful
verification of some of our results. We are greatly indebted
to Dezso Varga for many inspiring
remarks and for his precious help in the early stages of this work.
We gratefully acknowledge Krzysztof Golec-Biernat and Jacek Turnau
for discussion on some problems related to this study.\\
 This work was supported by the Polish State Committee for Scientific
Research under grant no. 1 P03B 097 29.

\vskip 1cm


\end{document}